\documentstyle[rotate,psfig,paspconf,11pt]{article}

\newcommand{\bc}{\begin{center}}
\newcommand{\ec}{\end{center}}
\newcommand{\bi}{\begin{itemize}}
\newcommand{\ei}{\end{itemize}}
\newcommand{\bd}{\begin{description}}
\newcommand{\ed}{\end{description}}
\def\sprop{$S_{\nu} \propto \nu^{\alpha}$}

\def\deg{\ifmmode $\setbox0=\hbox{$^{\circ}$}$^{\,\circ}
          \else    \setbox0=\hbox{$^{\circ}$}$^{\,\circ}$\fi\,}
\def\pdeg{\ifmmode $\setbox0=\hbox{$^{\circ}$}\rlap{\hskip.11\wd0 .}$^{\circ}
          \else \setbox0=\hbox{$^{\circ}$}\rlap{\hskip.11\wd0 .}$^{\circ}$\fi~}
\def\arcs{\ifmmode {^{\scriptscriptstyle\prime\prime}}
          \else $^{\scriptscriptstyle\prime\prime}$\fi~}
\def\arcm{\ifmmode {^{\scriptscriptstyle\prime}}
          \else $^{\scriptscriptstyle\prime}$\fi~}

\begin{document}

\include{cover3mm}

\bc
{\large \bf High Angular Resolution Monitoring of Prominent AGN at 86\,GHz}\\
\ec

\bc
T.P. Krichbaum, A. Witzel, and J.A. Zensus  \\

\vskip 0.25cm
\footnotesize{
Max-Planck-Institut f\"ur Radioastronomie, Bonn, Germany\\
}
\ec

\vskip 0.5cm\leftline{\bf Introduction}
\vskip 0.25cm

Synchrotron-self absorption in the inner regions of AGN-jets at cm-wavelengths and a stronger 
source activity (variability of flux density and jet structure)
at mm-wavelengths are two of the main motivations to study AGN with 
VLBI at mm-wavelengths. While VLBI imaging at 1 and 2\,mm is not yet
feasible (see Doeleman \& Krichbaum, this conference), 3\,mm-VLBI now provides
high angular resolution images with a dynamic range better than 100:1 for the brightest 
sources. 
With the growing number of antennas participating in global CMVA (\underline{C}oordinated 
\underline{M}illimeter \underline{V}LBI \underline{A}rray) experiments, more detailed
studies of compact radio sources, which were imaged previously only with small arrays (3-5 stations), are possible.
Here we present some new results from our ongoing 3\,mm-VLBI monitoring of 3C\,273 and 
3C\,454.3\footnote{A joint project of scientists at MPIfR, IRAM and Haystack}.

\vskip 0.5cm\leftline{\bf Results for 3C\,273}
\vskip 0.25cm

We observed 3C\,273 in January 1994 (4 stations), March 1995 (5 stations, including for the 
first, time Plateau de Bure), January 1996
(5 stations, including for the first, time Pie Town), and April 1997 (10 stations, 
including Sest). The maps of 1994 and 1995 show a
south-west oriented core-jet structure of $\sim 2$\,mas length, with at least 
two components receeding at superluminal speeds from the core (Krichbaum et al.
1997 \& 1998). VLBA observations of 3C\,273 at 22 and 43\,GHz by Marscher et al.
in 1995.15 is sufficiently close to our observation in 1995.18 to allow a detailed
comparison. In Figure 1 both maps are displayed. Dashed lines connect
the location of individual jet components, obtained from Gaussian
model fitting. The astoundingly good agreement of the component positions relative to the core
in both maps demonstrates (i) the reliability of the 3\,mm map, which results from
a much smaller VLBI array than the 7\,mm map, and (ii) allows an estimate of opacity effects and spectrum along 
the inner jet.
The core and the component located at $r \simeq 1.5$\,mas exhibit inverted spectra 
($\alpha_{\small \rm 43/86 GHz} = 0.4 \ldots 0.7$, \sprop), whereas the spectra of
the other components are steep ($\alpha_{\small \rm 43/86 GHz} = -0.3 \ldots -1.0$, Figure 1, right panel). The relative
offsets of the positions of the components at the two frequencies are small,
typically $\leq 0.1$\,mas. This is consistent with an optically thin jet with
only small opacity position shifts in the individual components.

The new maps obtained in 1996 and 1997 are shown in Figure 2. The increased
number of participating antennas has resulted in the improved uv-coverage and 
dynamic range of the images (dynamic range $\geq 300:1$).
The 1996 map indicated, for the first time, the existence of faint jet emission
beyond 2\,mas core-separation. This emission is confirmed and better visible in
the map of 1997, which shows faint emission even beyond the map area displayed here. 
The limited closure information for the short uv-spacings and remaining calibration 
uncertainties may cast some doubt on the reality of 
all details visible in the jet on mas-scales. However, we believe
that the basic jet structure is represented correctly. 
To check this, we superimpose a nearly simultaneously
observed 15\,GHz map (from the 2\,cm survey, Kellermann et al., 1998) on the 3\,mm map
in Figure 2 (right panel).

The emission at 15 and 86\,GHz track each other well. For the inner 3--4\,mas jet, the
86\,GHz emission is located at the center of the jet seen in the 15\,GHz map, 
indicating a central spine or ridgeline
with a smaller transverse width at higher frequencies (center brightening). At larger core separations,
both the 15 and 86\,GHz emission are displaced more to the south. At 86\,GHz this bend is
stronger than at 15\,GHz. This suggests edge brightening, with an opacity shift such, that
the 86\,GHz emission is closer to the jet boundary than the emission at 15\,GHz.

\begin{figure}[t]
\label{7a3mm}
\psfig{figure=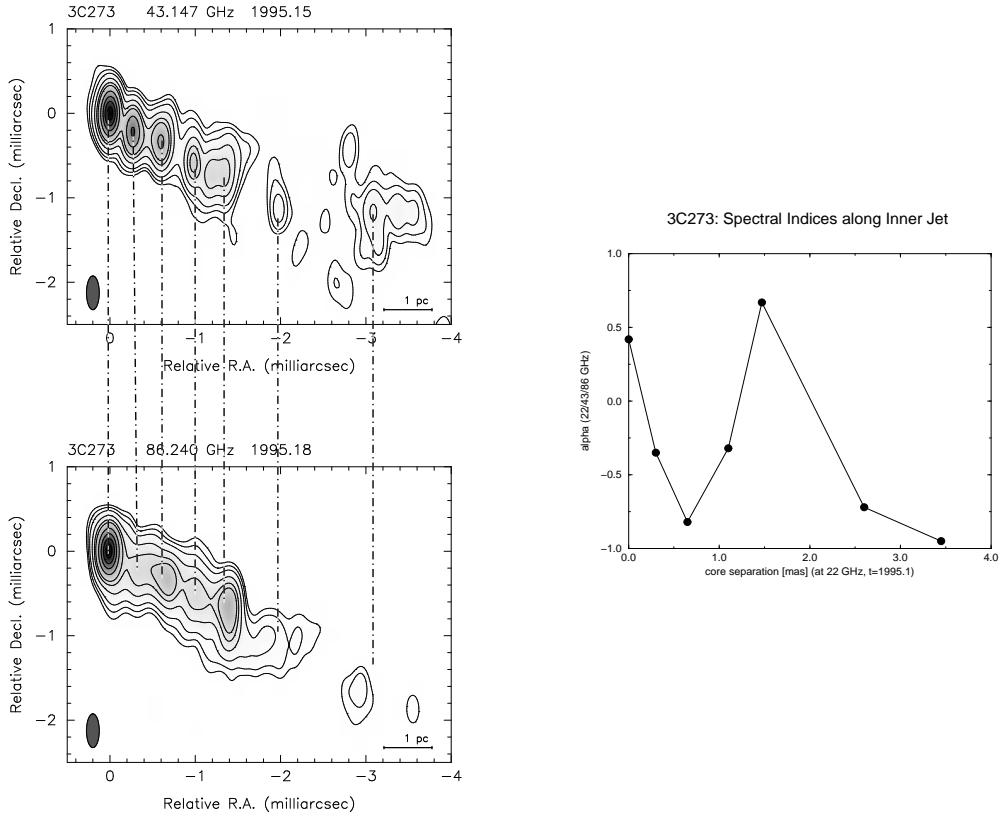,width=15cm}
\vspace{-0.5cm}
\caption{\small {\small \bf Left:} 3C\,273 at 43\,GHz (top left) and 86\,GHz (bottom left). Contour levels are 
-0.5, 0.5, 1, 2, 5, 10, 15, 30, 50, 70, and 90\,\% of the peak flux density of 5.4\,Jy/beam (top)
and 4.7\,Jy/beam (bottom). The restoring beam size for both maps is 
$0.4 ~{\rm x}~ 0.15$\,mas, oriented at pa$=0 \deg$. The data for the 43\,GHz map were kindly provided
by Alan Marscher.
{\small \bf  Right:} Spectral index (\sprop) plotted versus core separation for the components marked by dashed lines 
on the left.
} 
\end{figure}

\begin{figure}[t]
\label{273new}
\psfig{figure=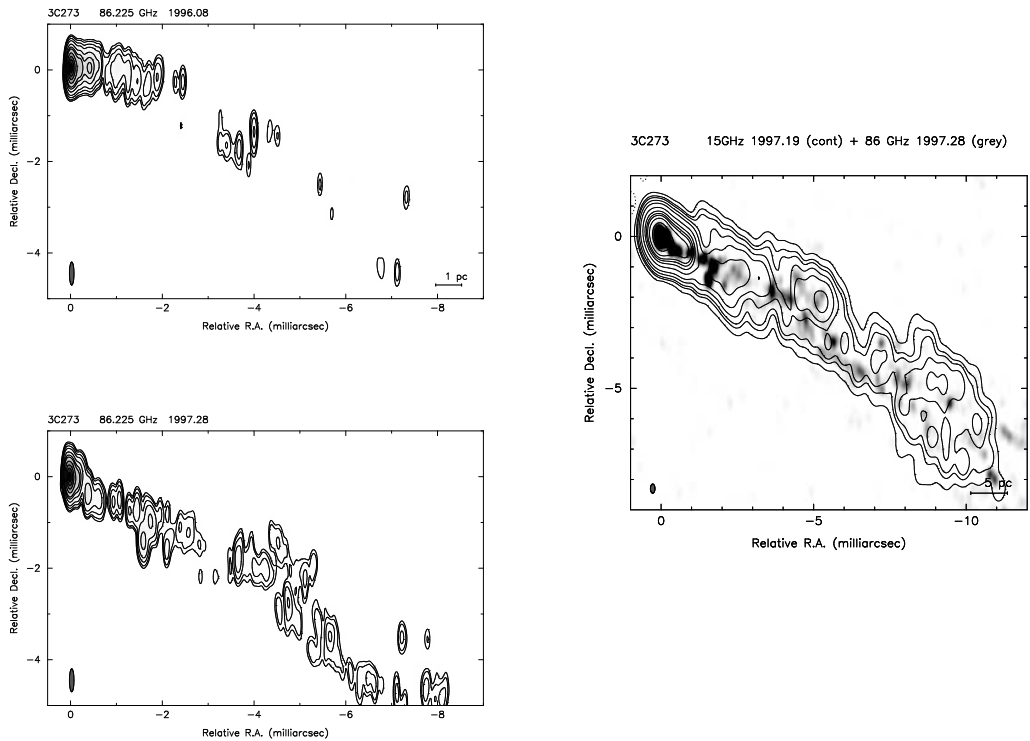,width=16cm}
\vspace{-0.5cm}
\caption{\small {\small \bf Left:} The mas-structure of 3C\,273 at 86\,GHz in 1996.08 (top) and 1997.28 (bottom). 
Contour levels are -0.3, 0.3, 0.5, 1, 2, 5, 10, 15, 30, 50, 70, and 90\,\% of the peak flux density of
7.3\,Jy/beam (top) and 7.5\,Jy/beam (bottom). The restoring beam size for both maps is
$0.5 ~{\rm x}~ 0.1$\,mas, oriented at pa$=0 \deg$.
{\small \bf Right:} a superposition of the 3C\,273 map of 1997.28 (bottom left) and a
nearly simultaneously observed 15\,GHz map is shown. Grey scale denotes 86\,GHz, contours
refer to 15\,GHz. The contour levels are
-0.1, 0.1, 0.3, 0.5, 1, 2, 5, 10, 15, 20, 30, 50, 70, and  90\,\% of the peak flux density
of 8.0\,Jy/beam. For the 15\,GHz map a restoring beam of
$0.9 ~{\rm x}~ 0.4$\,mas, oriented at pa$=0 \deg$, was used. 
}
\end{figure}

\vskip 0.5cm\leftline{\bf Results for 3C454.3}
\vskip 0.25cm

The OVV quasar 3C\,454.3 was observed in 1996 and 1997, in the same experiments as 3C\,273.
Previous 3\,mm maps showed an east-west oriented and slightly bent core-jet structure with 
two superluminally moving ($\sim 6$\,c) components embedded in a more diffuse underlying jet 
(Krichbaum et al., 1996 \& 1997). Near the core, the apparent motion is slower than 
at larger core separations (10--22\,c observed on a 2--7\,mas scale), but follows the 
general velocity gradient (cf.\ Pauliny-Toth, 1998).
The new 3\,mm-maps are shown in Figure 3 (left panel). Whereas the map of 1996 (top)
shows a relatively straight jet of $2-3$\,mas length, the map of 1997 (bottom) shows 
jet curvature to the south and then, beyond $r=1.5-2.0$\,mas, bending back towards 
north. We note that a sinusoidal jet path was suggested by Pauliny-Toth as a possible
explanation of the observed velocity variations in the outer mas-jet.
The change of jet orientation near 2\,mas is also seen in a 15\,GHz map,
observed in 1997.19 (Kellermann et al., 1998). A superposition of this map with the 86\,GHz map
is shown in Figure 3 (right panel). It is seen that the overall positional agreement between the
jet at 15 and 86\,GHz is not as good as for 3C\,273 (see Figure 2). This may be due to different
opacity effects in the two sources. We note that in 3C\,454.3
the largest offsets between the two frequencies appear at a core separation of $2-3$\,mas. 
It is possible that opacity effects could cause frequency dependent offsets between the mean
jet axis at 15 and 86\,GHz. These offsets would depend on the viewing angle and could become more 
pronounced in regions of
stronger jet curvature. The complexity of the source structure in 3C\,454.3, the still limited
uv-coverage (3C\,454.3 was observed during one day, 3C\,273 was observed on 3 consecutive days
with redundancy), and remaining calibration uncertainties, however, require further observations 
in order to confirm this effect.

\begin{figure}[t]
\label{454maps}
\psfig{figure=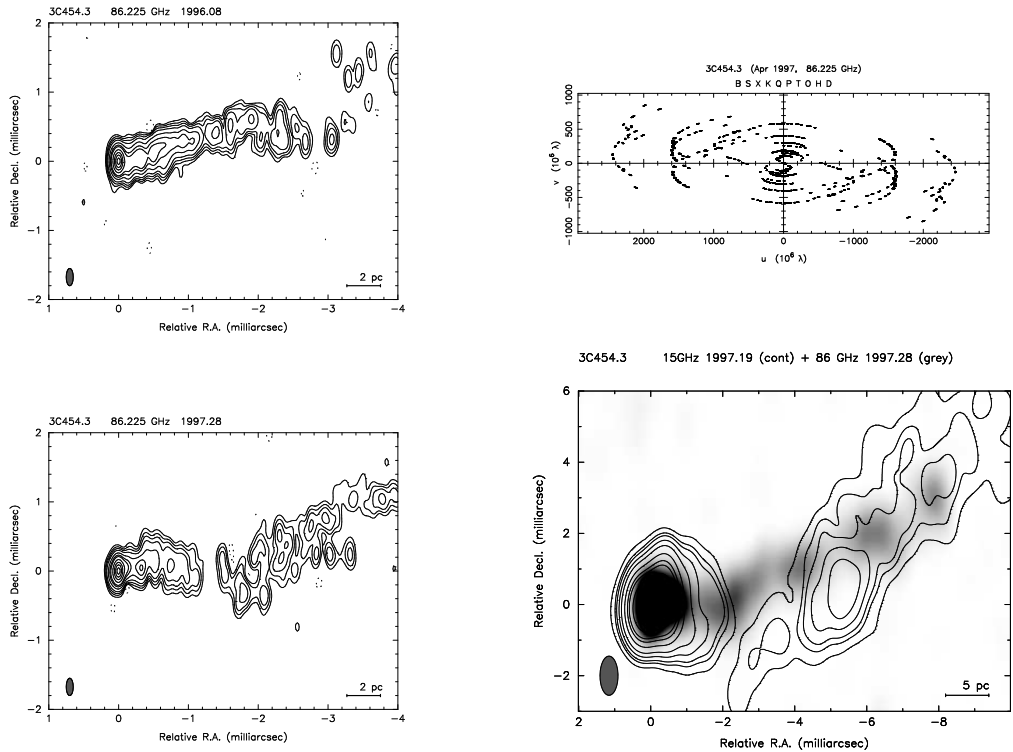,width=16cm}
\vspace{-0.5cm}
\caption{\small {\small \bf Left:} 3C\,454.3 at 86\,GHz observed in 1996.08 (top) and 1997.28 (bottom).
Contour levels in both maps are -0.2, 0.2, 0.5, 1, 2, 5, 10, 15, 30, 50, 70, and 90\,\%
of the peak flux of 0.9\,Jy/beam (top) and 1.2\,Jy/beam (bottom).
The restoring beam size is $0.25 ~{\rm x}~ 0.1$\,mas, oriented at pa$=0 \deg$.
{\small \bf Right:} The uv-coverage at 86\,GHz for 1997.28 (top). Below the
superposition of two nearly simultaneously observed
maps at 15\,GHz (contours) and at 86\,GHz (grey scale) at epoch 1997.2.
Contour levels are -0.1, 0.1, 0.3, 0.5, 1, 2, 5, 10, 15, 30, 50, 70, and 90\,\% of
the peak flux density of 3.9\,Jy/beam at 15\,GHz. Both maps are
convolved with a beam of $1.1 ~{\rm x}~ 0.5$\,mas size, oriented at pa$=0 \deg$.
}
\end{figure}

\vskip 0.5cm\leftline{\bf Conclusion}
\vskip 0.25cm

With up to 12 antennas participating in global CMVA experiments, maps with dynamic ranges of
300 to 500:1 can be made. For sources with long jets, like 3C\,273 and 3C454.3, the 
jet emission at relative large core separations is detectable, 
if short uv-spacings (100-1000\,km) are available.
Since at 3\,mm the beam size is small ($0.04-0.1$\,mas),
relatively wide fields have to be mapped ($>50-100$ beam sizes).
For a better imaging of these complex structures in the future,
a better uv-coverage on short and intermediate uv-spacings is needed. With regard to this, 
the participation of the `central' antennas of the VLBA (PT, LA, FD, NL),
and the short uv-spacings from the baseline Haystack--Quabbin and the European subarray
(Metsahovi--Onsala--Effelsberg, Pico Veleta--Plateau de Bure) will be of particular importance. 
The addition of the phased interferometers (BIMA, OVRO, Plateau de Bure) within the next
few years will help to improve the sensitivity of the whole array by at least a factor of 2--3.
With an expected single-baseline detection sensitivity of $50-100$\,mJy on baselines
to and between the phased interferometers (present threshold: $> 250$\,mJy), the number of
observable sources and the quality of images should increase dramatically.

\vskip 0.5cm\leftline{\bf References}
\begin{verse}
{\footnotesize
\vskip -0.5cm
Kellermann, K.I., Vermeulen, R.C., Zensus, J.A., Cohen, M.H., 1998, AJ, 115, 1295.\\

Krichbaum, T.P., Alef, W., Witzel, A., `The sub-parsec scale jets of AGN', 1996, 
in:  Extragalactic Radio Sources, IAU Symposium No.\ 175, eds. C. Fanti et al., 
Kluwer, Dordrecht, p. 11-13.\\

Krichbaum, T.P., Witzel, A., Graham, D., Lobanov, A.P., 1997,
`MM-VLBI Monitoring of Borad-Band Active Blazars', in: MM-VLBI
Science Workshop, eds. R. Barvainis and R.B. Phillips, Haystack MIT Cambridge, p. 3--9.\\

Krichbaum, T.P., Kraus, A., Otterbein, K., Britzen, S., and Witzel, A., 1998,
`Sub-mas Jets in Gamma-Active Blazars: Results from High Frequency VLBI',
in: IAU Colloquium No. 164: Radio Emission from Galactic and Extragalactic Compact
Sources, Astronomical Society of the Pacific Conference Series, Volume
144, eds. J.A. Zensus, G.B. Taylor, \& J.M. Wrobel, p.~37--38.\\

Pauliny-Toth, I.I.K., `Structural Variations in the Quasar 3C 454.3', 1998,
in: IAU Colloquium No. 164: Radio Emission from Galactic and Extragalactic Compact
Sources, Astronomical Society of the Pacific Conference Series, Volume
144, eds. J.A. Zensus, G.B. Taylor, \& J.M. Wrobel, p.~76--76.\\
}
\end{verse}
\end{document}